\documentclass[amsmath,amssymb,preprint,nofootinbib,aps]{revtex4-1}         

\usepackage{amssymb}
\usepackage{subfigure}
\usepackage{multirow}
\usepackage{amsfonts}
\usepackage{amsmath}
\usepackage{amssymb}
\usepackage{graphicx}
\usepackage{dcolumn}
\usepackage{times}
\usepackage{color}

\begin{document}

\title{Wavelet based detection of scaling behaviour in noisy experimental data}

\author{Y.~F. Contoyiannis}
\email[]{yiaconto@uniwa.gr}
\author{S. Potirakis}
\email[]{spoti@uniwa.gr}
\author{F.~K. Diakonos}
\email[]{fdiakono@phys.uoa.gr}

\affiliation{Department of Electrical and Electronics Engineering, University of West Attica, GR-12244, Egaleo, Greece}

\affiliation{Department of Electrical and Electronics Engineering, University of West Attica, GR-12244, Egaleo, Greece}

\affiliation{Department of Physics, University of Athens, GR-15784 Athens, Greece}
\date{\today}

\begin{abstract}
The detection of power-laws in real data is a demanding task for several reasons. The two, more frequently met, being: (i) real data possess noise which affects significantly the power-law tails and (ii) there is no solid tool for the discrimination between a power-law, valid in a specific range of scales, from other functional forms like log-normal or stretched exponential distributions. In the present report we demonstrate, employing simulated and real data, that using wavelets it is possible to overcome both of the above mentioned difficulties and achieve a secure detection of a power-law and an accurate estimation of the associated exponent. 
\end{abstract}
\pacs{}
\maketitle

\section{Introduction}

Scaling relationships of the form:
\begin{equation}
f(x)=c x^{-p} ~~~\mathrm{for}~\Delta > x > x_0
\label{eq:eq1}
\end{equation}
characterized as power-laws with characteristic exponent $p$, valid between the scales $x_0$ and $\Delta$, are wide spread in the analysis of signals occurring in measuring processes. They apply usually as a statistical property related to the tail of a distribution ($x \gg x_0$), describing the "weight" of the value $x$ in a measurement of a physical property $X$. Specific examples offer the power-spectrum $S(f)$ of a signal with $f$ the underlying frequency, encountered in the ${1 \over f}$ noise phenomenon, the statistical distribution $N(\tau)$ of the waiting times $\tau$ between beats in a heartbeat timeseries, the distribution of avalance sizes $s$ in self-organized criticality (SOC), the distribution of laminar lengths in the (spatial and temporal) order parameter fluctuations in critical systems etc. \cite{Power_laws}. In experimental signals, where the statistics are strictly limited, it is often the case that the tail of a calculated distribution is affected by random fluctuations, overriding the expected signal. Thus, although theoretically the presence of the power-law is expected to hold in the tail of a distribution, in the experimental observation it is restricted to the body of the distribution. This leads to enhanced uncertainty for the characterization of a specific distribution as a power-law. In addition, in physical systems, due to the finite size, it is often the case that the condition $\Delta \gg x_0$ necessary for a clear signature of the power-law behaviour becomes a simple inequality ($\Delta > x_0$) so that the range of validity of eq.~(\ref{eq:eq1}) is significantly shrank. Thus, there is only a small part of the related distribution which follows a power-law. Since the region where power-law applies is usually not strictly known, this affects significantly the estimate of the exponent $p$. Furthermore, in this case other functional forms like lognormal or stretched exponential distributions may describe equally well the recorded data, particularly if both factors (noisy tails, finite-size effects) are present \cite{Clauset2009}.

In this work we will demonstrate that the difficulties in the detection of a power-law in experimental data can be overcome using wavelets. The idea to apply wavelets for the detection of power-laws is not new. There is an extensive literature on this subject \cite{Powerlaws_wavelets}. The common point of view is to use wavelets for the analysis of the time-series on which the calculation of an emerging distribution, supposed to follow a power-law, is based. Priority in such a search is the recognition of self-similar patterns in the considered time-series. Here we will use the wavelets in a later phase, once the targeted distribution is already determined. We will show that the information for the appearance of a power-law is contained already in the lower scale coefficients which turn out to be insensitive to the mixture with noise. Even more, we demonstrate that one can locate zones in which scaling behaviour applies and the calculation of the associated scaling exponent is safely performed, even in the presence of noise, using these low-scale wavelet coefficients. 

In fact, our treatment inverses the problem concerning the search for power-law behaviour in experimental data. Instead of fitting with a power-law and applying suitable tests to the fitting results, here we consider the imprint of power-law behaviour on properties of the low-scale wavelet coefficients and we test the appearance of these properties in the statistical distributions obtained from the measured data. There are two significant advantages of the proposed method: (i) the robustness against the admixture of noise and (ii) the accurate determination of windows of scales in which the power-law description applies. 

In the following, we first extract the wavelet-coefficient properties we will use as a benchmark for the power-law appearance. These properties form the theoretical platform for the subsequent analysis. Then we will apply the developed criteria to a noise infected power-law distribution, for various noise amplitudes, demonstrating the robustness of the proposed approach. Next we apply the proposed method to simulated data and in particular to the distribution of the waiting times in the neighbourhood of zero (stable fixed point of the effective potential) in a 3-d Ising magnetization time series at the corresponding (pseudo)critical temperature. This distribution attains a power-law form as shown in \cite{Contoyiannis2002}. Finally, we will show the practical use of the proposed scheme applying it to a variety of distributions originating from experimentally determined time series.

\section{Power-law induced constraints to wavelet coefficients}

We start our discussion with the presentation of some properties of the wavelets which will be relevant for the following analysis. The translation and scaling properties of the mother wavelet in its discrete version are given by:
\begin{equation}
\psi_{j,k}(x)=2^{\frac{j}{2}} \psi(2^j x -k)~~~;~~~j,~k \in \mathbb{Z}
\label{eq:eq2}
\end{equation}
where the scaling is in powers of $2$. An arbitrary function $f(x)$ can be expanded in such a wavelet basis as follows:
\begin{equation}
f(x)=\displaystyle{\sum_{j=-\infty}^{\infty}}\displaystyle{\sum_{k=-\infty}^{\infty}} d_{j,k} \psi_{j,k}(x)
\label{eq:eq3}
\end{equation}
with:
\begin{equation}
d_{j,k}=\displaystyle{\int_{-\infty}^{\infty}}~dx \psi_{j,k}(x) f(x)
\label{eq:eq4}
\end{equation}
In the following we will consider a power-law of the form, valid for a finite system:
\begin{equation}
f(x)=\left\{ \begin{array}{ll} 0 \phantom{aaai}, & x  \notin [x_0,\Delta) \\ c x^{-p}~, &  \Delta > x \geq x_0 \\ \end{array} \right.
\label{eq:eq5}
\end{equation}
with $p \in \mathbb{R}$. Since $f(x)$ has finite support the sums in Eq.~(\ref{eq:eq3}) are restricted to $j \geq 0$ and $k \geq 0$.
For $p < 1$ one can further assume $x_0=0$ since the function (\ref{eq:eq5}) is integrable around $0$ in this case. Then $f(x)$
possesses the scaling property:
\begin{equation}
f(\frac{x}{b})=b^p f(x)
\label{eq:eq6}
\end{equation}
which, combined with Eq.~(\ref{eq:eq2}) holding for the mother wavelet, leads to:
\begin{equation}
d_{j',k'}=\frac{1}{\sqrt{b}}\displaystyle{\int_{-\infty}^{\infty}}~dx
2^{{j \over 2}} \psi(2^j {x \over b} -k) b^{-p} f({x \over b})
\label{eq:eq7}
\end{equation}
with $j'=j-\frac{\ln b}{\ln 2}$ and $k'=k$. Changing variable $x'={x \over b}$ we find:
\begin{equation}
d_{j',k'}=b^{-p+\frac{1}{2}} \displaystyle{\int_{-\infty}^{\infty}}~dx' \psi_{j,k}(x') f(x')=b^{-p+\frac{1}{2}} d_{j,k}
\label{eq:eq8}
\end{equation}
as the scaling property of the wavelet coefficients when $f(x)$ is of the form (\ref{eq:eq5}) with $p < 1$. Particularly, using $j'=j+1$, which means $b={1 \over 2}$, we find:
\begin{equation}
d_{j+1,k}=2^{p-\frac{1}{2}} d_{j,k}
\label{eq:eq9}
\end{equation}
Defining the ratio $R_{j,k}=\frac{d_{j,k}}{d_{j+1,k}}$ we find:
\begin{equation}
R_{j,k}=2^{{1 \over 2} - p}
\label{eq:eq10}
\end{equation}
which is independent of $j$ and $k$. Thus an inherent property of the wavelet decomposition of an ideal power-law is that:
\begin{equation}
\lambda=\frac{R_{j,k}}{R_{j+1,k}}=1~~~~~\mathrm{for}~\mathrm{all}~~j,~k
\label{eq:eq11}
\end{equation} 
Our proposal is to use this property for the characterization of a power-law obtained from a noisy experimentally observed signal. We argue that at the coarse scales (small $j$) the influence of the high frequency noise is suppressed and the relation (\ref{eq:eq11}) can filter the power-law behaviour. This will be clearly demonstrated in the following by calculating $\lambda$ for several sets of simulated and experimentally measured data. The notion of "ideal power-law" indicated above includes the constraints $p < 1$ and $x_0=0$. It is still needed to clarify how the relation (\ref{eq:eq11}) is modified when $p > 1$ and $x_0 \neq 0$. Notice that the condition $p > 1$ necessarily implies that $x_0 >0$ for $f(x)$ in Eqs.~(\ref{eq:eq1},\ref{eq:eq5}) representing a distribution. Additionally, it should be checked the modification of Eq.~(\ref{eq:eq11}) when $f(x)$ is replaced by a discrete power-law distribution of the form $f(i) \sim i^{-p}$ for $i=1,~2,..$. Such a situation will frequently occur in the examples considered in the following. Then one can write:
\begin{equation}
f(x)=c \displaystyle{\sum_{i=1}^{\infty}} \delta(x-i) x^{-p} 
\label{eq:eq12}
\end{equation}
and the integral in Eq.~(\ref{eq:eq4}) leads to the sum:
\begin{equation}
d_{j,k}=c \displaystyle{\sum_{i=i_{min}}^{i_{max}}} \psi_{j,k}(i) i^{-p} 
\label{eq:eq13}
\end{equation}
where $i_{min}$ and $i_{max}$ are determined by the support of $\psi_{j,k}(x)$, i.e. the $x$-interval for which $\psi_{j,k}(x)$ is different from zero. Both these scenarios will be explored in the next section.

\section{Haar analysis of power-law distributions}

To proceed we will continue our study using Haar wavelets for the expansion of a power-law function of the form in Eq.~(\ref{eq:eq5}), assuming $x_0 >0$ and $p >1$, or in Eq.~(\ref{eq:eq12}). The mother wavelet with non-vanishing values in the interval $(0,\Delta]$ is given in this case as:
\begin{equation}
\psi_H(x)=\theta({\Delta \over 2}-x) \theta(x)-\theta(x-{\Delta \over 2}) \theta(\Delta-x)
\label{eq:eq14}
\end{equation}
leading to the Haar wavelet basis:
\begin{eqnarray}
\psi_{j,k}(x)&=&\theta(k{\Delta \over 2^j}+{\Delta \over 2^{j+1}}-x) \theta(x-k{\Delta \over 2^j}) \nonumber \\
&& -\theta(x-k{\Delta \over 2^j}-{\Delta \over 2^{j+1}}) \theta((k+1){\Delta \over 2^j}-x)
\label{eq:eq15}
\end{eqnarray}
with $\theta(z)$ the Heaviside step function. Employing Eq.~(\ref{eq:eq4}) it is straightforward to calculate the coefficients $d_{j,k}$ for the expansion of $f(x)$, given in Eq.~(\ref{eq:eq5}), in the Haar basis. We start our calculations for the case $p < 1$ and $x_0=0$ which is discussed in a more general framework in the previous section. We find:
\begin{align}
d_{j,k}=\frac{c}{1-p}\sqrt{\frac{2^j}{\Delta}} \left[2 (k{\Delta \over 2^j}+{\Delta \over 2^{j+1}})^{1-p} - (k{\Delta \over 2^j})^{1-p} \right. \nonumber \\
\left. -  ((k+1){\Delta \over 2^j})^{1-p} \right]
\label{eq:eq16}
\end{align}
Clearly, for $p < 1$ and $x_0=0$ the coefficients $d_{j,k}$ obey the general scaling relations in Eqs.~(\ref{eq:eq9},\ref{eq:eq10},\ref{eq:eq11}) as expected. As a next step we calculate the coefficients
$d_{j,k}$ and the ratios $R_{j,k}$, $\lambda$ for $p > 1$ and $x_0 > 0$.
Obviously, for  $x_0 < k{\Delta \over 2^{j+1}}$ the expression in Eq.~(\ref{eq:eq16}) is still valid and therefore the scaling laws (\ref{eq:eq9},\ref{eq:eq10},\ref{eq:eq11}) are obeyed. In fact, whenever the condition $x_0 \ll \Delta$ is satisfied it exists a maximum $j_{max}$
for which the constraint:
\begin{equation}
\frac{x_0}{\Delta} < \frac{k}{2^{j+1}} 
\label{eq:eq17}
\end{equation}
is fulfilled. Then, provided that $k \geq 1$, Eq.~(\ref{eq:eq9}) is valid for $j < j_{max}-1$, Eq.~(\ref{eq:eq10}) is valid for $j < j_{max}-2$ while Eq.~(\ref{eq:eq11}) is valid for $j < j_{max}-3$. Since $j \geq 0$ we find that $j_{max} \geq 3$ which leads to the constraint $\Delta \geq 16 x_0$, i.e. the scaling law must hold for at least one and a half decade, a reasonable range for physical systems. Notice that for the applicability of our approach, we need to consider only the coarse grained scales (small $j$) which are less sensitive to the presence noise, thus the existence of a single set of $\{j,j+1,j+2\}$ fulfilling the condition (\ref{eq:eq17}), i.e. $j=0$, is sufficient. In such a case
we have that $\lambda=\frac{d_{0,0} d_{2,0}}{d_{1,0}^2}=1$ is the relevant condition for the appearance of a power-law between the scales $x_0$ and $\Delta$. Unfortunately, a complication occurs for $k=0$ since the condition (\ref{eq:eq17}) is violated. Thus, it needs more effort to handle this case. We will focus on this in the following.

For $k=0$ the Haar wavelet coefficients, when $x_0 > 0$, become:
\begin{align}
d_{j,0}=\frac{c}{1-p}\sqrt{\frac{2^j}{\Delta}} \left[2 ({\Delta \over 2^{j+1}})^{1-p} - x_0^{1-p} - ({\Delta \over 2^j})^{1-p} \right]
\label{eq:eq18}
\end{align}
leading to:
\begin{align}
R_{j,0}=2^{{1\over 2}-p} \left[ \frac{2 ({\Delta \over 2^{j+1}})^{1-p} - x_0^{1-p} - ({\Delta \over 2^j})^{1-p}}{2 ({\Delta \over 2^{j+1}})^{1-p} - 2 x_0^{1-p} - ({\Delta \over 2^j})^{1-p}} \right]
\label{eq:eq19}
\end{align}
Eq.~(\ref{eq:eq19}) after some algebraic manipulations becomes:
\begin{align}
R_{j,0}=2^{{1\over 2}-p} \left[ \frac{1 - {1 \over 2^p - 1} ({\Delta \over x_0 2^j})^{p-1}}{1 - {2 \over 2^p - 1} ({\Delta \over x_0 2^j})^{p-1}} \right]
\label{eq:eq20}
\end{align}
which now depends on $j$. However, for $\Delta \gg x_0$ and low values of $j$ the term in the brackets is close to ${1 \over 2}$ and consequently
the relation:
\begin{equation}
\lambda=\frac{R_{j,0}}{R_{j+1,0}} \approx 1
\label{eq:eq21}
\end{equation}
is still valid. 

Let us now consider the discrete power-law case described in Eq.~(\ref{eq:eq12}). Using the Haar basis one can easily extract the general form of the coefficients $d_{j,k}$ as follows:
\begin{align}
d_{j,k}=c \sqrt{\frac{2^j}{\Delta}} \left( \displaystyle{\sum_{i=max([k{\Delta \over 2^j}],1)}^{[k{\Delta \over 2^j}+{\Delta \over 2^{j+1}}]}} i^{-p} - \displaystyle{\sum_{i=[k{\Delta \over 2^j}+{\Delta \over 2^{j+1}}]+1}^{[(k+1){\Delta \over 2^j}]}} i^{-p} \right)
\label{eq:eq22}
\end{align}
where $[z]$ means the integer part of $z$. First we consider the case $k > 0$. Then, for $\Delta \gg 1$ and low values of $j$, in which we are interested, it holds $max([k{\Delta \over 2^j}],1)=[k{\Delta \over 2^j}]$. Equation (\ref{eq:eq22}) can be written as:
\begin{align}
d_{j,k}=c \sqrt{\frac{2^j}{\Delta}} \left( \zeta(p,[k{\Delta \over 2^j}])
+ \zeta(p,[(k+1){\Delta \over 2^j}]) \right. \nonumber \\ \left. -2 \zeta(p,[k{\Delta \over 2^j} + {\Delta \over 2^{j+1}}]+1) \right)
\label{eq:eq23}
\end{align}
where $\zeta(p,s)$ is the Hurwitz zeta function. For $j \leq j_{max}$ with $j_{max}$ such that $\frac{\Delta}{2^{j_{max}}} \gg 1$ we can use the asymptotic expansion of $\zeta(p,s)$ for $s \to \infty$:
\begin{equation}
\zeta(p,s) \stackrel{s \to \infty}{\approx} \frac{1}{2} s^{-p} + \frac{s^{1-p}}{p-1}
\label{eq:eq24}
\end{equation}
to rewrite $d_{j,k}$ as:
\begin{align}
d_{j,k} \approx {c \over p-1} \left(\frac{2^j}{\Delta}\right)^{p - {1 \over 2}} \left(k^{1-p} - (k+1)^{1-p} \right. \nonumber \\ \left. + \frac{(p-1)2^j}{2 \Delta} 
(k^{-p}-(k+1)^{-p}) \right)
\label{eq:eq25}
\end{align}
Since $\frac{2^j}{2 \Delta} \ll 1$ Eq.~(\ref{eq:eq25}) simplifies to:
\begin{equation}
d_{j,k} \approx {c \over p-1} \left(\frac{2^j}{\Delta}\right)^{p - {1 \over 2}}
\left( k^{1-p} - (k+1)^{1-p} \right)
\label{eq:eq26}
\end{equation}
and therefore 
\begin{equation} 
R_{j,k} \approx 2^{{1 \over 2}-p} \Rightarrow \lambda \approx 1
\label{eq:eq27}
\end{equation}
When $k=0$ we find:
\begin{equation}
d_{j,0}=c \sqrt{\frac{2^j}{\Delta}} \left( \zeta(p,1) + \zeta(p,[{\Delta \over 2^j}])-2 \zeta(p,[{\Delta \over 2^{j+1}}]+1) \right)
\label{eq:eq28}
\end{equation}
and employing the asymptotic expansion of $\zeta(p,s)$ for $s \gg 1$ we find:
\begin{equation}
d_{j,0} \approx c \sqrt{\frac{2^j}{\Delta}} \zeta(p)
\label{eq:eq29}
\end{equation}
which in turn leads again to $\lambda \approx 1$. Thus, we have shown that for a power-law the quantity $\lambda$ approaches $1$ for $\Delta \to \infty$ in a range of scales which necessarily includes the lowest $j$-values. We will show in the following that this property characterizes in a unique manner the presence of power-law behaviour in a general data set, constituting the backbone of the search method proposed in the present work.  

\section{Applying the wavelet protocol to data}

Based on the analysis of the previous section we present here an algorithm which can be directly applied to experimentally observed signals, searching for the presence of scaling behaviour. The proposed algorithm comprises the following steps:
\begin{itemize}
\item The numerical estimation of the ratio $\lambda $, based on Eq.~(\ref{eq:eq21}) for $j=0,k=0$, obtained using the formula:

\begin{equation}
\lambda =\frac{\frac{{{d}_{00}}}{{{d}_{10}}}}{\frac{{{d}_{10}}}{{{d}_{20}}}}=\frac{{{d}_{00}}{{d}_{20}}}{d_{10}^{2}}=\frac{\left( \displaystyle{\mathop{\sum }_{i=1}^{\frac{\Delta }{2}}}g\left( i \right)-\displaystyle{\mathop{\sum }_{\frac{\text{ }\!\!\Delta\!\!\text{ }}{2}}^{\Delta }}g\left( i \right) \right)\left(\displaystyle{\mathop{\sum }_{i=1}^{\frac{\Delta }{8}}}g\left( i \right)-\displaystyle{\mathop{\sum }_{i=\frac{\Delta }{8}}^{\frac{\Delta }{4}}}g\left( i \right)\right)}{{{\left( \displaystyle{\mathop{\sum }_{i=1}^{\frac{\Delta }{4}}}g\left( i \right)-\displaystyle{\mathop{\sum }_{i=\frac{\Delta }{4}}^{\frac{\Delta }{2}}}g\left( i \right) \right)}^{2}}}
\label{eq:eq30}
\end{equation}

where $g\left( i \right),~~i=1,2,\ldots $, denotes the signal values, and 
$8<\Delta <{{\Delta }_{max}}$ (${{\Delta }_{max}}$ the length of signal). According Eq.~(\ref{eq:eq30}), we expect that, when the signal possesses a scaling behaviour, $\lambda$ will approach the value $\lambda \approx 1$ for sufficiently large $\Delta$.

\item We denote as ${{I}_{\lambda }}(\epsilon_{\lambda})$ the interval of $\Delta$-values within which $\lambda$  converges to $1$ with a prescribed accuracy $\epsilon_{\lambda}$. For this interval we define the ratio $R$ through the following formula (using again $j=0,k=0$):

\begin{equation}
R=\frac{{{d}_{00}}}{{{d}_{10}}}=\frac{1}{\sqrt{2}}\frac{\left( \displaystyle{\mathop{\sum }_{i=1}^{\frac{\Delta }{2}}}g\left( i \right)-\displaystyle{\mathop{\sum }_{\frac{\Delta }{2}}^{\Delta }}g\left( i \right) \right)}{\left( \displaystyle{\mathop{\sum }_{i=1}^{\frac{\Delta }{4}}}g\left( i \right)-\displaystyle{\mathop{\sum }_{i=\frac{\Delta }{4}}^{\frac{\Delta }{2}}}g\left( i \right) \right)}
\label{eq:eq31}
\end{equation}

\item We locate the interval ${{I}_{R}}\subseteq {{I}_{\lambda }}$ in $\Delta$-space, for which the $R$-values stabilize around a mean value $<R>$ with accuracy $\epsilon_R$ and we calculate this mean value.

\item For the estimation of the $p-$exponent we use the discrete version of the test function in Eq. (\ref{eq:eq5}) $f\left( i \right)=c{{i}^{-p}},~~~i=1,2,3,4,\ldots$ and we solve numerically the equation (\ref{eq:eq31}) for the given $<R>$-value with respect to $p$. Note that when the signal is particularly noisy one may observe dispersed $\lambda$-values close to $1$. In such a case, the scaling behaviour is only approximate and the noise generates exponential tails which suggest the use of a test function $h\left( i \right)=c{{i}^{-p}}{{e}^{-qi}},~~~i=1,2,3,4,\ldots $ in Eq.~(\ref{eq:eq31}) for the description of the noisy signal. As in the pure power-law case we solve Eq.~(\ref{eq:eq31}) for the pair $(p,q)$ with the condition $q \ll 1$ which guarantees that the power-law, at least approximately, holds, allowing the use of Eq.~(\ref{eq:eq31}). In practice, the additional parameter $q$, introduced to estimate the divergence from scaling behaviour, leads to a self-consistent description for $q-$values in the interval $(0.01-0.1)$ (empirical result).

\end{itemize}

To demonstrate how this algorithm works in practice, we provide a series of examples, starting from simulated data and extending up to experimentally measured signals.

The simulated data are generated as a sequence of uncorrelated random numbers sampled from a specific distribution $P(r)$. Let us denote by $\{r_n\}$ (with $n=1,2, \dots N$) such a timeseries. We consider three cases: (i) exponentially, (ii) log-normal and (iii) power-law distributed $r_n$. In all cases the length of the corresponding timeseries is $N=150000$. Noise is added in two ways into the considered timeseries: (a) multiplicatively, i.e. by transforming at each step $r_n$ to $\tilde{r}_n=r_n (1 + a \xi_n)$ or (b) additively, with the transform $\hat{r}_n=r_n + a \xi_n$  with $\xi_n$ uniformly distributed in $[0,1)$. We use the term amplitude noise for the case (a) and the term additive noise for (b).

In Fig.~1a we show the function $\lambda(\Delta)$ in the noiseless scenario comparing the result for a power-law distribution with exponent $p=1.67$ (red circles) to the result corresponding to a slow exponential decay (exponent $0.05$, blue circles) and to that of a log-normal distribution with variance $\sigma=1$ (green circles). We observe that, for the power-law, $\lambda(\Delta)$ converges quite rapidly to $1$ and it is clearly distinguished form the other two cases. Notice that, since the timeseries is finite, the distribution of the $r_n$-values can become exactly zero for finite $r$ and the $\lambda(\Delta)$ jumps trivially to $1$ as it can be easily shown using Eq.~(\ref{eq:eq30}). This happens for the log-normal distribution. To avoid this effect, the $\Delta_{max}$ value is taken to be smaller than the value $r$ leading first to $P(r)=0$ in each case. In Fig.~1b we show the functions $\lambda(\Delta)$ for the three cases (exponential, log-normal, power-law) presented in Fig.~1a, now with the inclusion of amplitude noise with $a=0.9$. We observe that the time-series with $\{r_n \}$ originating from a power-distribution, is practically unaffected by the presence of noise, in contrast to the exponentially generated $r_n$. The timeseries generated by $r_n$ distributed according to the log-normal distribution is affected less than the exponential case. A remarkable difference is that, in the presence of noise, the distribution $P(r)$ for $r_n$ containing a log-normal component is non-vanishing for a wider range of $\Delta$-values. This is clearly seen is Fig.~1c where we plot the distributions $P(r)$ for the three considered cases in the presence of amplitude noise with $a=0.9$. Finally, in Fig.~1d we show the behaviour of the function $R(\Delta)$ for the timeseries containing a power-law component in the noiseless (black line) as well as in the noisy (amplitude noise with $a=0.9$) case (red line) we observe the saturation of $R(\Delta)$ for $\Delta \gg 8$. Setting the asymptotic value of $R$ in Eq.~(\ref{eq:eq31}) we solve for $p$ finding $R=1.68$ for both the noiseless and the noisy case. Due to finite statistics a small value of $q$ (here $q \approx 0.01$) is required for the description of the simulated data \footnote{The required $q$-value decreases with increasing statistics and decreasing at the same time the size of the bin $\Delta r$ used in the calculation of $P(r)$.}. 

\begin{figure}[tbp]
\centering
\includegraphics[width=1.05\textwidth]{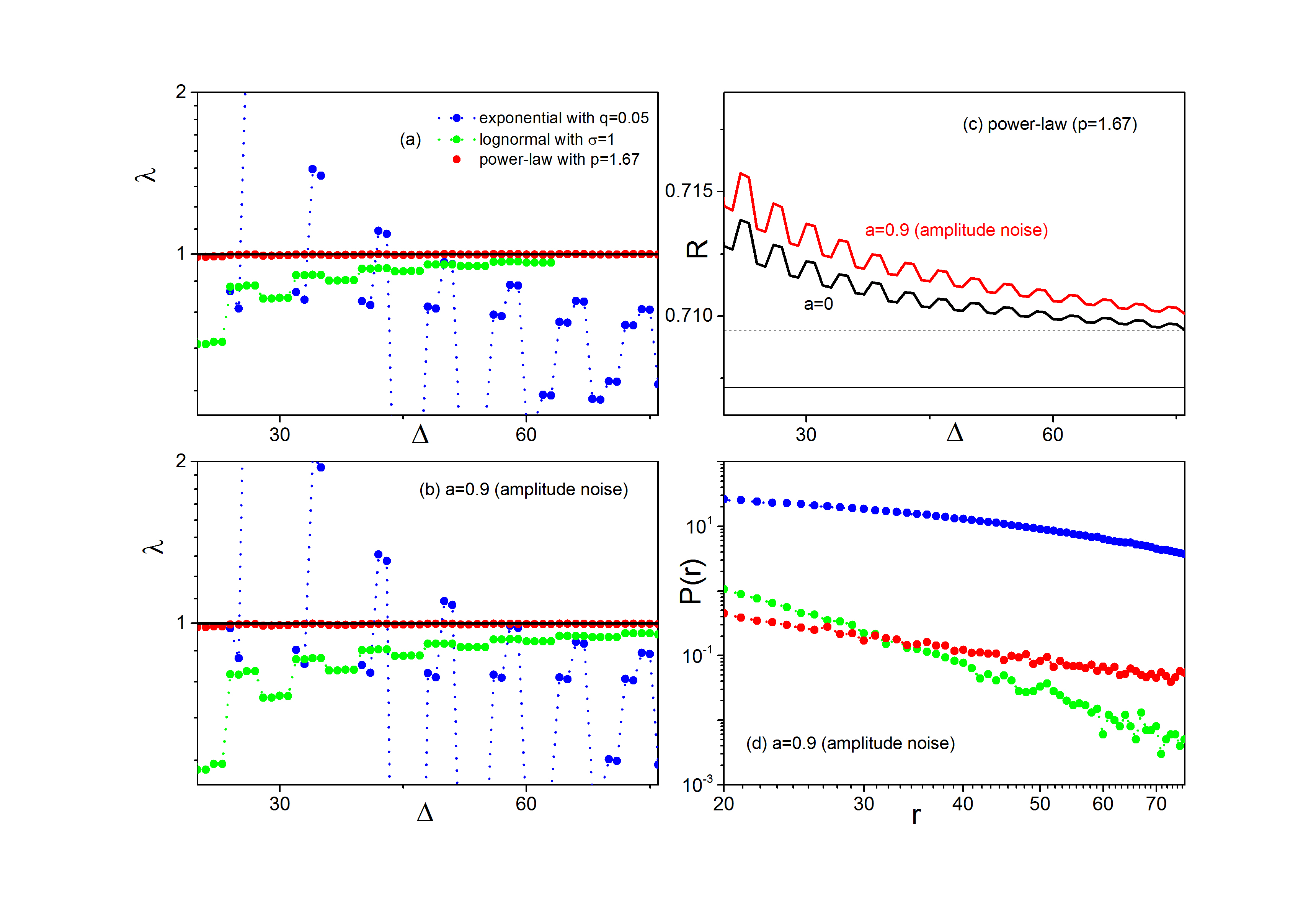}
\caption{{\it (a) The function $\lambda(\Delta)$ given in Eq.~(\ref{eq:eq30}), calculated from timeseries of random numbers distributed exponentially (blue circles), log-normal (green circles) and power-law (red circles). The black solid line at $\lambda=1$ is shown to guide the eye.  In (b) we show the functions $\lambda(\Delta)$ for similar timeseries obtained from the timeseries in (a) adding amplitude noise with $a=0.9$. In (c) we show the function $R(\Delta)$ for the power-law case in the noiseless (black line) and the noisy (red line) case. Both lines converge to $R \approx 0.71$ leading to $p=1.68$. Finally, in (d) we show the distributions of the timeseries values used in 
(a).}} 
\label{fig:fig1}
\end{figure} 

To explore further the insensitivity of $\lambda(\Delta)$ to noise effects in the case of power-law distributed $\{r_n\}$, we consider also the influence of strong additive noise with $a=5$. In such a case the $P(r)$ distributions are strongly disordered as one can see in Fig.~2a. No clear evidence of power-law behaviour is observed in the distribution of $r_n$ values containing a power-law component (red circles). Despite this, the corresponding function $\lambda(\Delta)$ still converges to $1$ for $\Delta > 80$, as it can be seen in Fig.~2b. We observe also that $\lambda(\Delta)$ for the timeseries with a log-normal component (green circles) approaches $1$, but the corresponding distribution $P(r)$ becomes zero before reaching it. Furthermore, Fig.~2c displays the remarkable stability of the function $R(\Delta)$ in the power-law case. We have considered also cases with even stronger additive noise and the observed robustness is sustained. However, with increasing $a$ the range of allowed $\Delta$-values shrinks and the distributions $P(r)$ vanish rapidly preventing the applicability of the proposed wavelet based algorithm for $a >10$.

\begin{figure}[tbp]
\centering
\includegraphics[width=0.65\textwidth]{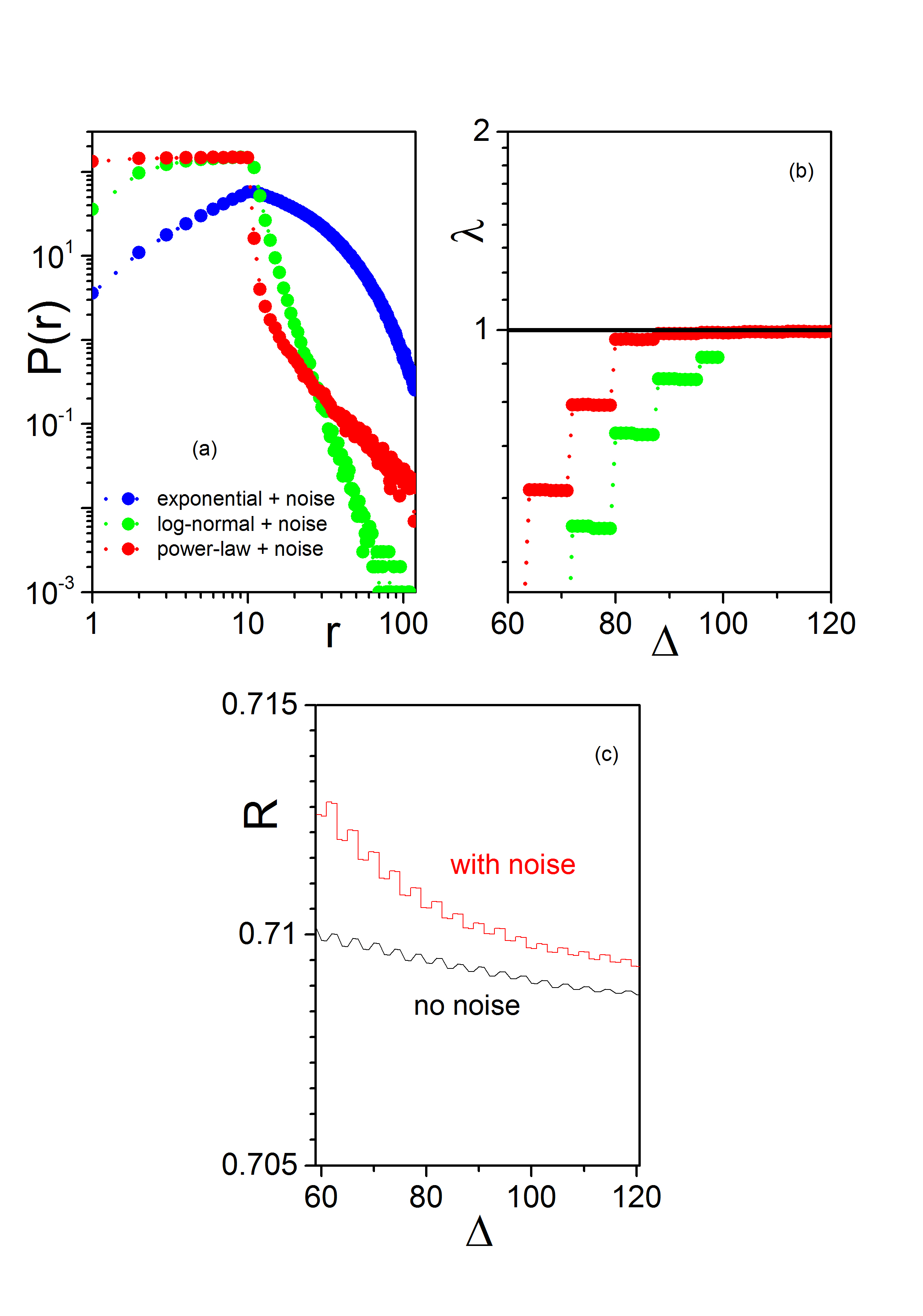}
\caption{{\it (a) The distributions of the timeseries values in the presence of additive noise with $a=5$ for exponentially (red circles), log-normal (green circles) and power-law (red circles) distributed $r_n$. In (b) we show the corresponding  $\lambda(\Delta)$-functions for the log-normal (green circles) and the power-law (red circles) case. Notice that the exponential does not fit at all in the presented window of $\lambda$. Finally, in (c) we show the function $R(\Delta)$ for the power-law case comparing the noiseless (black line) with the noise (red line) case.}}
\label{fig:fig2}
\end{figure} 

Continuing with the applications of our approach we present one more characteristic example allowing the detection of power-law behaviour in distributions obtained from simulated timeseries. It is obtained from the magnetization timeseries in the 3-D Ising model simulation with Metropolis algorithm. As shown in \cite{Contoyiannis2002}, the waiting times $\tau$ (measured in sweeps) in a narrow region around zero magnetization are power-law distributed with an exponent $p=1+{1 \over \delta}$, when $T=T_c$. For the 3-D Ising model $\delta \approx 5$ (isothermal critical exponent), while $T_c$ is the corresponding (pseudo)critical temperature for the ferromagnetic transition depending on the lattice size. For $T < T_c$ the waiting time distribution $P(\tau)$ attains exponential tail. In Fig.~3a we plot $P(\tau)$ for $T \approx T_c$ (red circles) and $T < T_c$ (blue circles) using simulated magnetization timeseries in a cubic Lattice with $L=20$. We observe a clear difference in the two distributions. This difference is clearly reflected also in the behaviour of the corresponding functions $\lambda(\Delta)$ shown in Fig.~3b. For the $T=4.545\approx T_c$ case the power-law behaviour is imprinted in the property $\lambda \approx 1$ for a wide range of $\Delta$ values while the deviation from criticality leads to the removal from the value $1$ for $T=4.45 < T_c$. Using the quantity $R$ in Eq.~(\ref{eq:eq31}) we find for $T\approx T_c$ the exponent $p=1.18$ which is very close to $1+{1 \over \delta}$ for the 3-D Ising universality class ($\delta \approx 5$).  

\begin{figure}[ht] 
  \subfigure[]{% 
    \includegraphics[width=.6\textwidth]{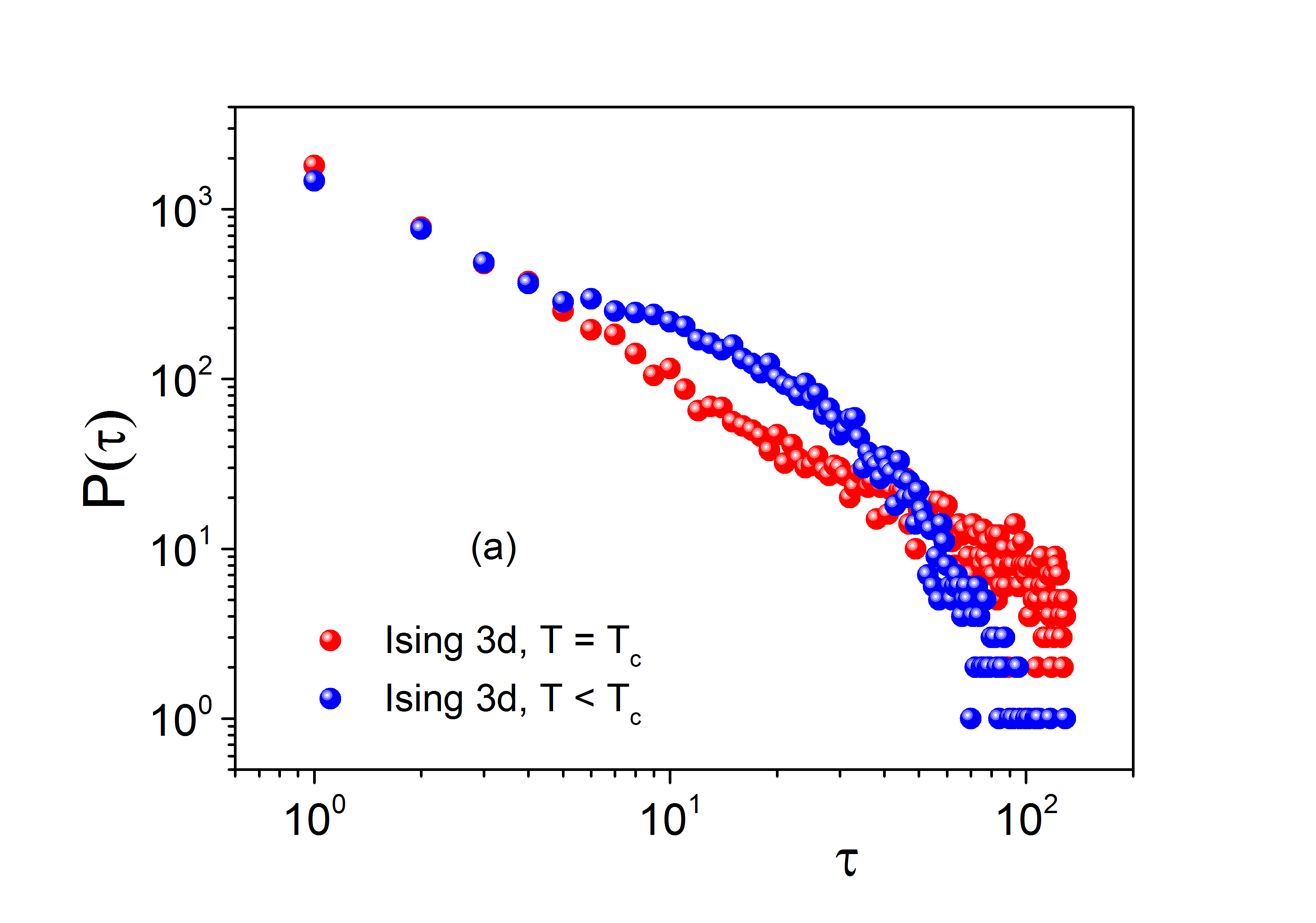} \label{fig:3a} 
  } 
  \quad 
  \subfigure[]{% 
    \includegraphics[width=.6\textwidth]{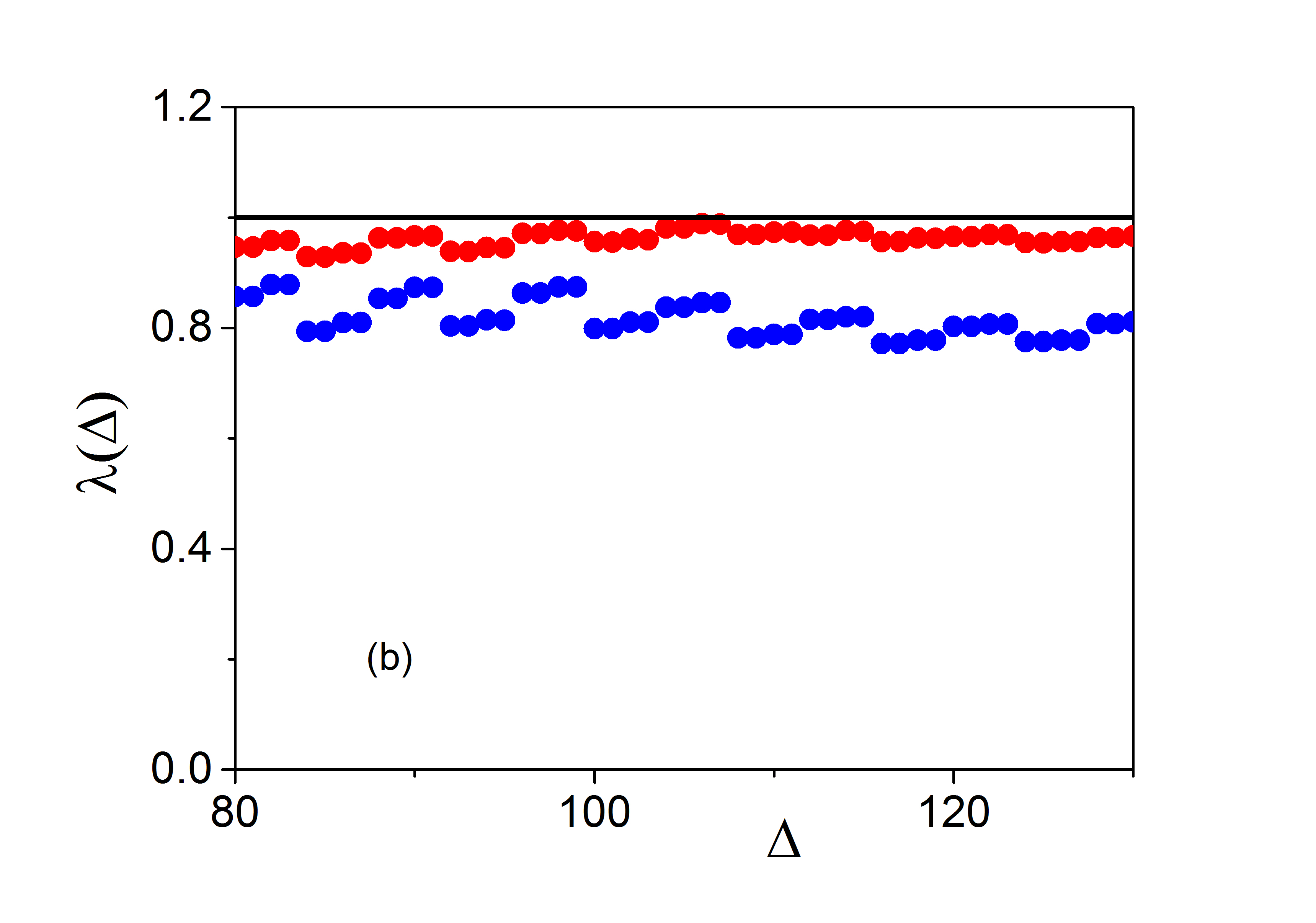} \label{fig:3b} 
  } 
  \caption{{\it In (a) we show the waiting time distribution $P(\tau)$ for the 3D-Ising model at $T=4.545$ (red circles, pseudocritical) and $T=4.45$ (blue circles). In (b) we plot the functions $\lambda(\Delta)$ for each case. }} 
\end{figure}

Before closing this section we present some examples demonstrating the application of the wavelet based power-law detection (WBPLD) method to experimentally measured timeseries. In all examples the waiting time distributions $P(\tau)$ around the most frequent value of the corresponding timeseries are calculated. In Figs.~4(a-c) we plot these distributions. In (a) we show the distribution obtained from ECG timeseries for human beings \cite{Contoyiannis2013}.  The red circles represent the case of healthy individuals while the blue circles correspond to individuals with infraction.  In (b) is displayed the waiting time distribution around the most frequent value of the membrane potential fluctuations of pyramidal neurons in the CA1 region of rat hippocampus \cite{Kosmidis2018}. Finally, in (c) is shown the waiting time distribution around the most frequent value of the preseismic E/M emission in MHz channel \cite{Contoyiannis2005}. In the panel in Figs.~4(d-f) we show the corresponding $\lambda(\Delta)$ functions. The approach to $1$ is evident supporting the presence of power-law behaviour in all three cases. The corresponding power-law exponents, calculated through $R$ in Eq.~(\ref{eq:eq31}), coincide with the values calculated in the literature with an accuracy $< 1\%$. Notice that in the plot in Fig.~4d the blue circles, which correspond to the case of humans with infraction $\lambda(\Delta)$, departs from the value $1$. This is in full agreement with the findings in \cite{Contoyiannis2013}. 
\begin{figure}[tbp]
\centering
\includegraphics[width=0.9\textwidth]{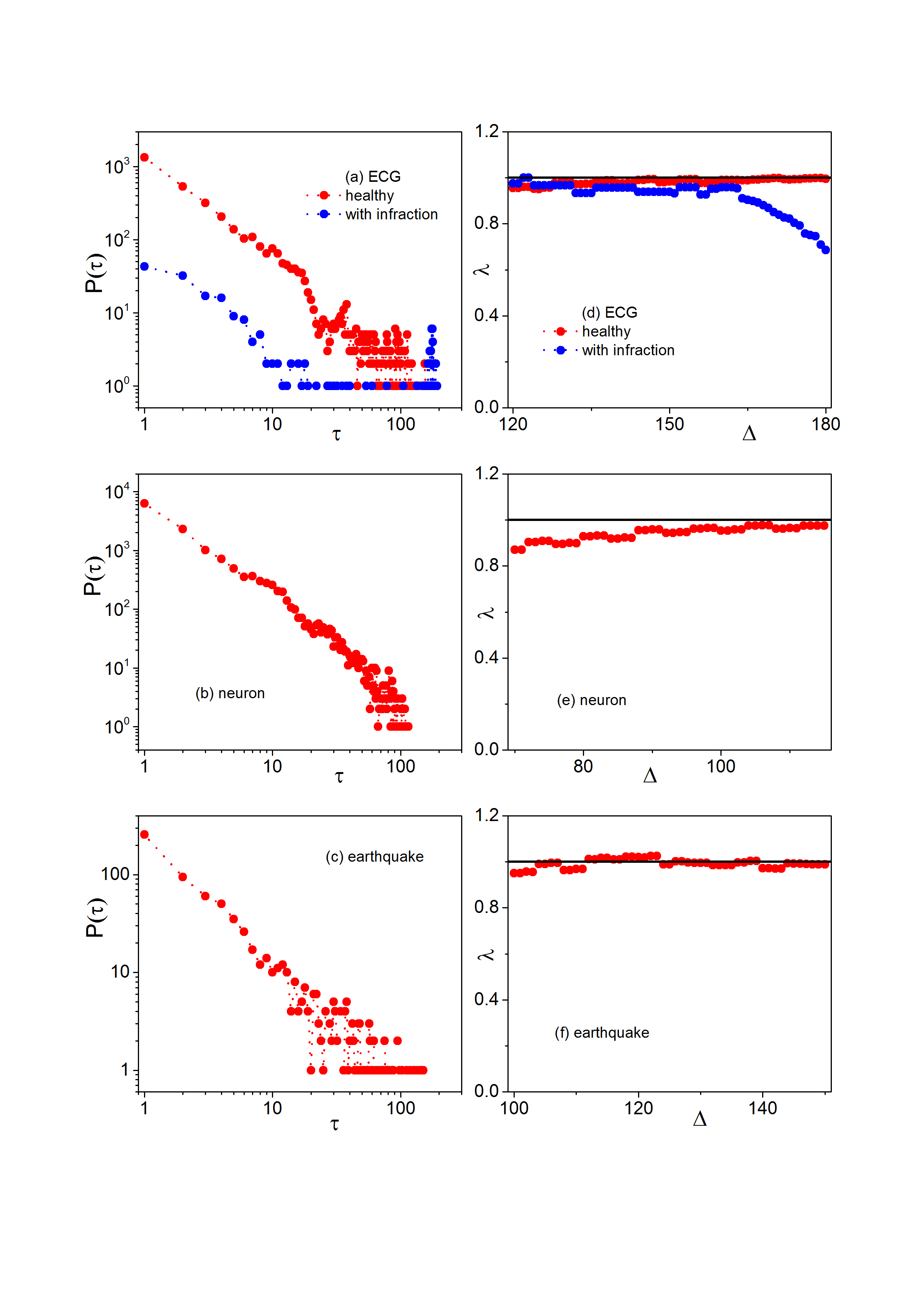}
\vspace*{-1cm}\caption{{\it The waiting time distribution $P(\tau)$ for: (a) human ECG timeseries of healthy individuals (red circles) and individuals with infraction (blue circles), (b) membrane potential fluctuations of pyramidal neurons in the CA1 region of rat hippocampus and (c) preseismic E/M emission in MHz channel. In (d), (e), (f) we show the corresponding functions $\lambda(\Delta)$. The black line at $\lambda=1$ is drawn to guide the eye.}}
\label{fig:fig4}
\end{figure} 

All the results of our analysis, including the estimation of the associated power-law exponents, are summarized in the following table: 
\begin{table}
\begin{center}
\begin{tabular}{| c | c | c | c | c |}
\hline 
Signal & $\Delta_{max}$ & $R$ & $p$ (fit) & $p$ (WBPLD) \\
\hline
\hline
3D Ising & 139 & 0.8082 &   1.18 & 1.18  \\ 
\hline
ECG human & 180 & 0.7305 &   1.32 & 1.32  \\
\hline
Neuron rat & 115 & 0.7496 & 1.55 & 1.53 \\
\hline
Earthquake & 150 & 0.7753 & 1.31 & 1.31 \\
\hline
%\hline
\end{tabular}
\end{center}
\caption{{\it $\Delta_{max}$, saturating $R$-value and power-law exponents $p$  obtained from the waiting time distributions in timeseries of simulated and experimentally recorded data. In fourth column we show also the result obtained from a power-law fit neglecting the tail of the distribution.}}
\end{table}

%  Healty Heart of Human [“The Earth as a living planet : Human-type
%diseasesin the earthquake preparation process” Y. F . Contoyiannis, S.M. Potirakis, K.
%Eftaxias, Nat. Hazards Earth Syst. Sci 13(2013) 125-139.]
%(c) real data : Neuron of mouse [“ Traits of criticality in membrane potential
%fluctuations of pyramidal neurons in the CA1 region of rat hippocampus “ Efstratios
%K. Kosmidis, Yiannis F. Contoyiannis , Costas Papatheodoropoulos ,Fotios K.
%Diakonos . European Journal of Neuroscience , https://doi.org/10.1111/ejn.14117
%(2018). ]
%(d) real data : Preseismic E/M emission in MHz channel “Monitoring of a preseismic
%phase from its electromagnetic precursors”. Y. Contoyiannis, P. Kapiris and K.
%Eftaxias. Phys. Rev. E 71, 1(2005).  

\section{Conclusions}

We have developed a computational tool for the efficient detection of power-law behaviour in distributions generated from experimentally recorded data. In contrast to the standard fitting procedures like least squares or maximum likelihood, which requires the adjustment of the parameters of the simulating function to optimally describe the observed distribution, in our method we first derived a number of properties characterizing the wavelet transform of the target distribution (power-law in our case) independently from the value of the associated exponent and subsequently we searched for the appearance of these properties in the experimental data. Having verified the presence of power-law behaviour, the corresponding exponent is obtained solving an algebraic equation. The great advantage of this procedure is that due to the scaling properties of the wavelet basis, it is possible to observe power-law behaviour which is valid only between two arbitrary scales which do not need to differ significantly. Furthermore, the occurrence of different scaling behaviours at different scales is also detectable within this framework. Finally, we have shown that the proposed scheme is capable to filter windows of power-law behaviour, allowing at the same time the safe estimation of the corresponding exponent, even in the presence of intensive noise. We have demonstrated the efficiency of our approach in a number of examples with increasing complexity. Thus, the proposed treatment introduces a novel strategy in the model description of experimentally observed data,  providing an alternative to the standard fitting procedures.

\end{document}